\documentclass[aps,singlecolumn,notitlepage,pra,superscriptaddress,longbibliography,nofootinbib]{revtex4-1}
\usepackage{amssymb}
\usepackage{amsmath}
\usepackage{graphicx}
\usepackage{color}
\usepackage[english]{babel}
\usepackage[latin1]{inputenc}

\definecolor{darkblue}{rgb}{0.,0.24,0.51}
\definecolor{britishracinggreen}{rgb}{0.0, 0.26, 0.15}
\definecolor{darkgreen}{rgb}{0,0.60,.2}

\usepackage{hyperref}
\hypersetup{%
   pdfpagemode=None, 
   pdfstartpage=1,
   pdfmenubar=true,
   pdftoolbar=true,
   colorlinks = true,
   linkcolor=blue,
   citecolor=blue,
   urlcolor=blue,
   bookmarksopen=false
 }

\newcommand{\edit}[1]{{\color{black}#1}}

\newcommand{\beq}{\begin{equation}}
\newcommand{\eeq}{\end{equation}}

\newcommand{\bra}[1]{\langle #1|}
\newcommand{\ket}[1]{|#1\rangle}
\newcommand{\mean}[1]{\langle #1 \rangle}
\newcommand{\vectgr}[1]{{\boldsymbol#1}}    

\newcommand{\D}{\mathcal{D}}
\newcommand{\SP}{\mathcal{S}}
\newcommand{\W}{\mathcal{W}}

\begin{document}

\title{Topological characterization of chiral models through their long time dynamics}
\author {Maria Maffei}
\email{maria.maffei@icfo.eu}
\affiliation{ICFO -- Institut de Ciencies Fotoniques, The Barcelona Institute of Science and Technology, 08860 Castelldefels (Barcelona), Spain}
\affiliation{Dipartimento di Fisica, Universit\'a di Napoli Federico II, Complesso Universitario di Monte Sant'Angelo, Via Cintia, 80126 Napoli, Italy}
\author {Alexandre Dauphin}
\affiliation{ICFO -- Institut de Ciencies Fotoniques, The Barcelona Institute of Science and Technology, 08860 Castelldefels (Barcelona), Spain}
\author{Filippo Cardano}
\affiliation{Dipartimento di Fisica, Universit\'a di Napoli Federico II, Complesso Universitario di Monte Sant'Angelo, Via Cintia, 80126 Napoli, Italy}
\author {Maciej Lewenstein}
\affiliation{ICFO -- Institut de Ciencies Fotoniques, The Barcelona Institute of Science and Technology, 08860 Castelldefels (Barcelona), Spain}
\affiliation{ICREA -- Instituci{\'o} Catalana de Recerca i Estudis Avan\c{c}ats, Pg.\ Lluis Companys 23, 08010 Barcelona, Spain}
\author {Pietro Massignan}
\affiliation{ICFO -- Institut de Ciencies Fotoniques, The Barcelona Institute of Science and Technology, 08860 Castelldefels (Barcelona), Spain}
\affiliation{Departament de F\'isica, Universitat Polit\`ecnica de Catalunya, Campus Nord B4-B5, 08034 Barcelona, Spain}
\date{\today}

\begin{abstract}
We study chiral models in one spatial dimension, both static and periodically driven. 
We demonstrate that their topological properties may be read out through the long time limit of a bulk observable, the mean chiral displacement. 
The derivation of this result is done in terms of spectral projectors, allowing for a detailed understanding of the physics.
We show that the proposed detection converges rapidly and it can be implemented in a wide class of chiral systems. Furthermore, it can measure arbitrary winding numbers and topological boundaries, \edit{it applies to all non-interacting systems, independently of their quantum statistics}, and it requires no additional elements, such as external fields, nor filled bands.
\end{abstract}
\maketitle

Topological phases of matter constitute a new paradigm by escaping the standard  Ginzburg-Landau theory of phase transitions.  These exotic phases appear without any symmetry breaking and are not characterized by a local order parameter, but rather by a global topological order. 
In the last decade, topological insulators have attracted much interest \cite{Hasan2010}. 
These systems are insulators in their bulk but exhibit current carrying edge states protected by the topology. 
A classification of topological insulators in terms of their discrete symmetries and their spatial dimensionality has been obtained in the celebrated periodic table of topological insulators and superconductors \cite{Chiu2016}. 
The topological invariant characterizing these models can be derived from the bulk Hamiltonian and allows one to recover the so called bulk-edge correspondence, namely that the number of topologically protected edge states is proportional to the topological invariant. 
A famous example of this correspondence can be found in the Quantum-Hall effect where the quantization of the Hall conductance is rooted in the current-carrying protected edge states \cite{Laughlin1981,Thouless1982,Halperin1982}. 
The ensemble of (natural and artificial) topological insulators is steadily growing, and these have been by now synthetically engineered in a multitude of physical systems such as atomic \cite{Atala2013,Stuhl2015,Mancini2015,Leder2016,Goldman2016, Meier2016}, superconducting \cite{Beenakker2016}, photonic \cite{Hafezi2013, Rechtsman2013, Zeuner2015, Lu2016, Mukherjee2017} and acoustic platforms~\cite{Xiao2015, Peano2015, Peng2016}.

This work focuses on one-dimensional (1D) topological insulators possessing chiral symmetry. As a consequence of the chiral symmetry, the different sites of the unit cell can always be regarded as part of two sublattices. The topological invariant of the bulk,  the winding number $\W$, allows one to predict the number of  zero energy edge states. 
1D chiral topological insulators have been realized in numerous platforms as ultracold atoms \cite{Atala2013, Meier2016}, photonic crystals \cite{Zeuner2015}, photonic quantum walks \cite{Kitagawa2012,Barkhofen2016, Cardano2016, Cardano2017,Flurin2017}. Let us notice that the 1D chiral Hamiltonian can be static or the effective Hamiltonian of a Floquet system. In the latter case, the topology can be richer than its static counterpart \cite{Asboth2012,Asboth2013,Asboth2014,Nathan2015}. 
In both cases, two different approaches to characterize the topology of such systems have been proposed and implemented experimentally. 
\edit{The first one is to look at intrinsic properties of the system. 
The second one is to measure the response to an external change. 
The observation of edge states \cite{Kitagawa2012, Mazza2015, Meier2016, Simon2017} and the measurement of the winding number through the mean chiral displacement \cite{Cardano2017} belong to the first category.
The measurement of the winding number by interferometric architectures \cite{Atala2013, Flurin2017}, by introducing losses \cite{Rudner2009, Zeuner2015, Rakovszky2017}, and by scattering measurements \cite{Barkhofen2016} belong to the second category.}

\edit{ Here we  generalize the notion of mean chiral displacement introduced in Ref.~\cite{Cardano2017}, and we present an intrinsic measurement of the topology for 1D chiral Hamiltonians with an arbitrary (even) number of sites per unit cell. This measure is based on the real-time evolution of an initially localized, single-particle state. Since this probes the free dynamics (i.e., without interactions), the method applies equally well to fermionic and bosonic systems, without requiring a specific band filling. Moreover, the measurement is carried out inside the bulk, so that the method also applies to systems with periodic boundary conditions (i.e., in a ring geometry). The method directly probes the Hamiltonian of interest, without resorting to additional external forces (which is generally required in 2D). Finally, this detection scheme even applies to a class of dissipative bosonic systems, as long as the losses act uniformly on all sites of the lattice.
Practically, our proposal only requires the single-site-resolved measurement of the density, and therefore it is implementable in a wide class of natural and artificial systems.} 

The plan of the paper is as follows. 
In Sec.~\ref{sec:chiralmod} we introduce the chiral models, define the relevant projectors, and present various equivalent definitions of the winding which may be used when either the Hamiltonian or its eigenstates are known.
In Sec.~\ref{sec:real-time-detection} we derive our main results, which permit the characterization of topology through the real time dynamics of the system.
We start by introducing various classes of localized states, we define the operators to characterize their displacement, and finally we show how the the real-time evolution of an initially localized state provides direct access to the chiral winding number.
In Sec.~\ref{sec:SSH4} we introduce the simplest 1D chiral model with internal dimension $\D=4$, we discuss its properties, and apply our findings to study its topology.
In Sec.~\ref{sec:Floquet_SSH_4} we show how our results may equivalently well be used to characterize periodically-driven systems; specifically, we consider the example of a quantum walk with four internal degrees of freedom. In Sec. \ref{sec:LR-SSH}, we consider an SSH model with additional staggered long-range hoppings and show that our detection method can measure windings greater than $1$.
In Sec. \ref{sec:outlook}, we discuss possible experimental implementations of these systems, and present outlook and conclusions.

\begin{figure}[t]
\begin{center}
\includegraphics[scale=1]{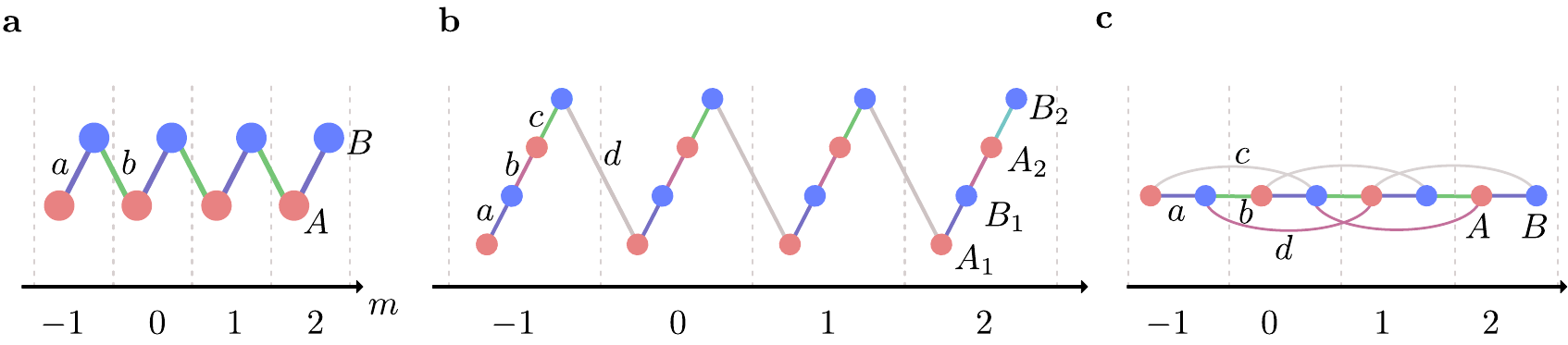}
\caption{\label{fig:sketch_models}
{\bf 1D chiral models}. Sketch of the SSH (\textbf{a}), the SSH$_4$ (\textbf{b}) and the SSH model with staggered long range hoppings  (\textbf{c}). Red and blue sites belong respectively to the A and B sublattices, thin lines denote hoppings, and the unit cells of the lattice are indexed by the integer $m$. The corresponding Hamiltonians are explicitly chiral, as these contain no term coupling a sublattice directly with itself.}
\end{center}
\end{figure}

\section{Chiral Models}\label{sec:chiralmod}
\subsection{Chiral Hamiltonians and spectral projectors}\label{subsec:projectors}

In this paper we consider 1D chiral systems, described by a tight-binding Hamiltonian on a lattice with $N$ unit cells, and $\mathcal{D}$ sites per unit cell. 
An Hamiltonian $H$ is said to possess chiral symmetry if there exists a local (i.e., acting only within a single unit cell), unitary and Hermitian operator $\Gamma$ (so that $\Gamma^2=\mathbb{I}$) which anticommutes with the Hamiltonian \cite{Asboth2016,Chiu2016}, 
\beq
\label{eq:cs}
\Gamma H\Gamma=-H.
\eeq
We will only consider the case where $\D$ is even, else chiral model necessarily present flat bands at zero energy.
Equation~\eqref{eq:cs} has a remarkable consequence: in a chiral system one may always identify two intertwined sublattices $A$ and $B$, of equal length, and the Hamiltonian ``swaps" them. 
Working in the  ``canonical basis" where the $\mathcal{D}/2$ first elements are in the sublattice $A$, the Hamiltonian has therefore a completely block-off-diagonal form
\beq \label{offDiagH}
H=\left(\begin{tabular}{cc}
0 & $h^\dagger$\\
$h$& 0
\end{tabular}\right),
\eeq
 and the chiral operator has diagonal form

 \beq
 \Gamma=\left(\begin{array}{cc}
\mathbb{I} & 0\\
0& -\mathbb{I}
\end{array}\right).
\eeq 

The simplest 1D chiral model is the one introduced by Su, Schrieffer and Heeger (SSH) to describe electrons hopping along polyacetylene chains \cite{Su1979}. These chains present a dimerized structure, and the dynamics of the electrons is described to a very good approximation by a tight-binding model with staggered tunnelings, so that the unit cell is composed of two sites (i.e., it has internal dimension $\D=2$). In this paper, we will be interested in describing more general chiral systems, characterized by $\D\geq2$; the simplest example of a chiral model with $\D=4$ is shown in Fig.~\ref{fig:sketch_models}b, and we will refer to it as the SSH$_4$ model. In the following, we will also consider a periodically driven $\D=4$ chiral model (see Fig.~\ref{fig:QW}a) and a model with winding greater than $1$, See Fig.~\ref{fig:sketch_models}c. Another example of chiral model with $\D>2$ may be found in Ref.\ \cite{Rakovszky2017}.

To further characterize chiral models, let us denote the eigenstates of the Bloch Hamiltonian by $\ket{\psi_j(k)}$, with $j=\pm1,\ldots,\pm \D/2$, and the corresponding energies by $E_j(k)$ (with $E_j(k)>0$ for $j>0$).  For simplicity, unless explicitly needed, we will often drop momentum indices. This notation is chosen to highlight the fact that eigenstates of chiral systems appear in chiral-partners pairs, $\Gamma\ket{\psi_j}=\ket{\psi_{-j}}$, with energies satisfying $E_{-j}=-E_j$.
Let us further introduce two classes of projectors, which will prove useful in the next sections: the projectors on energy eigenstates,
\beq
P_{j}=\ket{\psi_j}\bra{\psi_{j}},
\eeq
and the projectors on the eigenstates of positive energy, minus the one on the states of negative energy,
\beq
Q=\sum_{j=1}^{\D/2}Q_j=\sum_{j=1}^{\D/2}P_{j}-P_{-j}.
\eeq
These definitions agree with the usual ones of, e.g., Ref.\ \cite{Ryu2010}, since a fermionic system can be chiral only if its Fermi energy is set at $E_F=0$. The $Q$-matrix has the following properties  (see, e.g., \cite{Chiu2016}): (i) is Hermitian and unitary, (ii) it satisfies $Q^2=\mathbb{I}$, so that its eigenvalues are simply $\lambda_Q=\pm1$, (iii) \edit{it is diagonal in any basis of eigenvectors of $H$}, (iv) once expressed in the canonical basis, it becomes block-off-diagonal, 
\beq
Q=\left(\begin{tabular}{ll}
 0 & $q^\dagger$  \\
 $q$ & 0
\end{tabular}\right),\eeq
with $q$ unitary. Finally, the chiral operator may be written as a sum of ``partial chiral operators",
\beq
\Gamma=\sum_{j=1}^{\D/2}\Gamma_j=\sum_{j=1}^{\D/2}\ket{\psi_j}\bra{\psi_{-j}}+\ket{\psi_{-j}}\bra{\psi_{j}}.
\eeq

\subsection{The winding number $\W$}\label{subsec:winding}
One dimensional chiral models are characterized by a quantized invariant, the {\it winding number} $\W$. The latter is an integer, which may be positive or negative, 
The bulk-edge correspondence relates the winding of a periodic system to the number of protected edge states which appear when edges are inserted in the system. In particular, the number of edge states on each edge is exactly equal to $|\W|$.
\edit{The system's Zak phase $\gamma$ \cite{Zak1989} corresponds to $\pi\W$ (eventually, depending on the convention, modulo $2\pi$).}

The winding number $\W$  may be found in various equivalent ways, starting from either the Hamiltonian with periodic boundary conditions, or the associated $Q$-matrix, or their eigenstates. The first one is in terms of the winding of the lower-left block $h$ of the off-diagonal Hamiltonian \cite{Asboth2013}, 
\begin{align} \label{gammaAO}
\W=  \oint \frac{{\rm d}k}{2\pi i}  {\rm Tr}[h^{-1}\partial_k h]
= \oint \frac{{\rm d}k}{2 \pi i} \partial_k \log[{\rm Det}(h)]
= \oint\frac{{\rm d}k}{2 \pi} \partial_k \arg[{\rm Det}(h)]
= \sum_{j=1}^{\D/2}\oint \frac{{\rm d}k}{2 \pi i} \partial_k \log h_j,
\end{align}
where $\oint{\rm d}k=\int_{-\pi}^{\pi}{\rm d}k$ indicates an integral over the whole Brillouin zone, and $\{h_j\}$ denote the complex eigenvalues of the matrix $h$.
We have implicitly assumed that the Hamiltonian is gapped at zero energy, so that both $H$ and $h$ are invertible, and we have used the fact that the integral of the derivative of a continuous and periodic function is zero over a complete period. The winding of the model is therefore given by the cumulative winding of all eigenvalues of $h$ around the origin of the complex plane.

Equivalently, one may extract $\W$ from the winding of $q$, the lower-left block of the $Q$-matrix \cite{Ryu2010},
\begin{align}\label{gammaR}
\W= \oint \frac{{\rm d}k}{2 \pi i} {\rm Tr}[q^{-1}\partial_k q] 
=  \oint\frac{{\rm d}k}{2 \pi}\partial_k\arg[{\rm Det}(q)].
\end{align}
The last equality may be simply demonstrated writing $q=\sqrt{{\rm Det}(q)}u$, so that $u\in SU(2)$, and exploiting the fact that the winding of any $SU(2)$ matrix is zero

Alternatively, as discussed in Ref.\ \cite{Mondragon-Shem2014}, the winding may be computed through the integral over the Brillouin zone of the skew polarization $\SP=\sum_{j \in {\rm occ.}}\SP_{j}$,
\begin{align}\label{gammaM}
\W= \oint \frac{{\rm d}k}{\pi} \SP(k).
\end{align} 
The quantity $\SP_j = i \bra{\Gamma \psi_{j}}\psi_j'\rangle$ (with $\ket{\psi_j'}\equiv \partial_k\ket{\psi_j}$) may be shown to be a purely real number, and ${\rm occ.}$ denotes the set of occupied bands (i.e., of negative energies). 
From these definitions, it is clear that the winding is not a property of a single band but rather of the $\D/2$ negative (or positive) energy bands, which all contribute to its value.

\subsubsection{Winding of the SSH model}

To make a concrete example, for the usual SSH model we have $h=a+be^{ik}$, and $q=h/|h|$, so that $\arg h= \arg q$.
As $k$ traverses the Brillouin zone from 0 to $2\pi$, both complex numbers $h$ and $q$ wind once in the positive (counter-clockwise) direction, so that the winding is either 0 or $1$, depending on whether these circle enclose or not the origin. 
The normalized eigenvectors which are also chiral partners are $\ket{\psi_\pm}=\frac{1}{\sqrt2}\left(\frac{\sqrt{a^2+2 a b \cos (k)+b^2}}{a+b e^{i k}},\pm1\right)$, and $\oint \frac{{\rm d}k}{\pi} i\bra {\psi_+}\psi_-'\rangle$ equals either 0 or 1.
All methods above therefore coincide in dictating that the winding of the non-trivial SSH model is $\W=1$.

\section{Detection of topological invariants in real time}
\label{sec:real-time-detection}

We now proceed to illustrate the key finding of this work, i.e., that the winding number emerges in the long time limit of an observable, the ``mean chiral displacement", measured over initially localized states. This detection requires no precise knowledge about the Hamiltonian's details (apart from the fact that it is chiral symmetric), and it simply relies on the detection of the average position of the wavepacket (or of the single particle) within each sublattice.

To proceed, we start by introducing the chiral localized states, then we discuss displacement operators, and finally we present our main results.

\subsection{Chiral localized states}
A generic localized state on the site $m=0$ may be written as a superposition of Bloch eigenstates $\ket{\psi_{j}}$,
\begin{align}\label{PsiBar}
\overline{\ket{\Psi}}
\equiv \oint\frac{{\rm d}k}{\sqrt{2\pi}}\ket{\Psi}
= \oint\frac{{\rm d}k}{\sqrt{2\pi}}\sum_{j=\pm1,\ldots,\pm\D/2}\alpha_j\ket{\psi_{j}}.
\end{align}
In the following, localized states will be denoted with an overbar, and we will be interested in a particular subclass of these:
  the ``chiral localized states" $\overline{\ket{\Gamma_{j}}}$,
\begin{align}\label{Gamma_j}
\overline{\ket{\Gamma_{j}}}=\frac{{\rm sgn}(j)\overline{\ket{\psi_{j}}}+\overline{\ket{\psi_{-j}}}}{\sqrt{2}}=
\oint\frac{{\rm d}k}{\sqrt{2\pi}}\ket{\Gamma_{j}},
\end{align}
where
$\overline{\ket{\psi_{j}}}
= \oint\frac{{\rm d}k}{\sqrt{2\pi}}\ket{\psi_{j}}$. 
The chiral localized states are eigenstates of the partial chiral operator, $\Gamma_j\ket{\Gamma_{j'}}=\delta_{jj'}{\rm sgn}(j)\ket{\Gamma_j}$, and yield $\langle Q_j\rangle_{\Gamma_{j'}}=0$.

\subsection{Position and displacement operators}
We now introduce the position operator $\hat{m}$ (where the integers $m$ label whole unit cells, as shown in Fig.~\ref{fig:sketch_models}), and the ``chiral position" operator $\widehat{\Gamma m}\equiv\Gamma\hat{m}$. 
In the following, we will work in units where the length of a unit cell is set to unity.
The position operator in momentum space is represented as usual by a derivative,
\beq
\bra{k}\hat{m}\ket{\tilde{k}}
=\sum_{m,\tilde{m}}\bra{k}m\rangle\bra{m}\hat{m}\ket{\tilde{m}}\bra{\tilde{m}}\tilde{k}\rangle
=\sum_{m,\tilde{m}}m\delta(m-\tilde{m})\frac{e^{i(\tilde{k}\tilde{m}-km)}}{2\pi}
=i\partial_k\sum_{m}\frac{e^{i(\tilde{k}-k)m}}{2\pi}
=i\partial_k\delta(\tilde{k}-k).
\eeq
As an immediate consequence, one finds for example that the mean position of a generic localized state $\overline{\ket{\Psi}}$ as defined in of Eq.~\eqref{PsiBar} is obviously zero,
\beq
\langle\hat{m}\rangle_{\overline{\Psi}}
=\oint\frac{{\rm d}k{\rm d}\tilde{k}}{2\pi}\bra{\Psi}k\rangle\bra{k}\hat{m}\ket{\tilde{k}}\langle \tilde{k}\ket{\Psi}
=\oint\frac{{\rm d}k}{2\pi}\langle i\partial_k\rangle_{\Psi}
=\frac{i}{2} \oint\frac{{\rm d}k}{2\pi}\partial_k\bra{\Psi}\Psi\rangle=0.
\eeq
In the last step, we have used that $i\partial_k$ is a Hermitian operator.

Let us now consider the time evolution of $\overline{\ket{\Psi}}$. Its mean displacement after time $t$ is given by:
\begin{align}\label{MeanDisplacement}
\langle{\hat{m}}(t)\rangle_{\overline{\Psi}} = \oint \frac{{\rm d} k}{2\pi} \langle U^{-t} (i\partial_k) U^t\rangle_{\Psi},
\end{align}
\edit{where $U^t\equiv e^{-iHt}$ is the unitary evolution operator, and $U^{-t}\equiv e^{iHt}$ its inverse.}
The corresponding {\it mean chiral displacement} is:
\begin{align}\label{MeanChiralDisplacement}
\langle\widehat{\Gamma m}(t)\rangle_{\overline{\Psi}} \equiv \oint \frac{{\rm d} k}{2\pi} \langle U^{-t}\Gamma (i\partial_k) U^t\rangle_{\Psi}.
\end{align}
Finally, we define the {\it chiral average displacement}:
\begin{align}\label{MeanChiralDisplacementAlternative}
\langle\Gamma\cdot \hat{m}(t)\rangle_{\overline{\Psi}}
 \equiv \oint \frac{{\rm d} k}{2\pi} \langle \Gamma U^{-t} (i\partial_k) U^t\rangle_{\Psi}
 = \oint \frac{{\rm d} k}{2\pi} \langle U^t \Gamma (i\partial_k) U^t\rangle_{\Psi}.
\end{align}

\subsection{Measure of the winding number}
\label{eq:meas_winding}
The mean displacement at time $t$ of Eq.~\eqref{MeanDisplacement} can be written as (see App.~\ref{sec:app:meandisplacement} for details) 
\begin{align}\label{MeanDispOverlinePsi}
\langle \hat{m}(t)\rangle_{\overline{\Psi}} 
&=\sum_{j=1}^{\D/2}
\oint\frac{{\rm d}k}{2\pi} \Big\{
t\partial_kE_{j}\mean{Q_{j}}_\Psi 
+\SP_{j}\sin(2tE_{j}) \mean{i Q_{j}\Gamma_{j}}_\Psi
- \SP_{j}[1-\cos(2tE_{j})]\mean{\Gamma_{j}}_\Psi 
\Big\} \\\nonumber
&+\sum_{j, j'=\pm 1,\ldots,\D/2 \textrm{ and }|j|\neq|j'|}
\oint\frac{{\rm d}k}{2\pi} i\bra{\psi_j}\psi_{j'}'\rangle \langle \Psi\ket{\psi_j}\bra{\psi_{j'}}\Psi\rangle e^{i t (E_j-E_{j'})},
 \end{align}
where $Q_{j}$ and $\Gamma_{j}$ are the projectors introduced in Sec. \ref{subsec:projectors}, and $\SP_{j}$ is the skew polarization introduced in Sec. \ref{subsec:winding}. 
Equation~\eqref{MeanDispOverlinePsi} generalizes the one we found for the special case $\mathcal{D}=2$ in Ref.~\cite{Cardano2017}. In particular, the operator we had generically indicated with $\Gamma_\perp$ in our earlier work is now uniquely identified by the explicit expression $iQ_j\Gamma_j$. 
When evaluated on the chiral localized states  $\overline{\ket{\Gamma_j}}$, the mean displacement reduces to
\begin{align}
\langle \hat{m}(t)\rangle_{\overline{\Gamma_j}}
 &=-{\rm sgn}(j)\oint\frac{{\rm d}k}{2\pi}\SP_{j}[1-\cos(2tE_{j})].
\end{align}
We therefore find that the chiral average  displacement, $\langle \Gamma \cdot\hat{m}(t)\rangle$, when summed on the chiral localized states with $j > 0$, converges in the long-time limit to minus one half of the winding number $\mathcal{W}$,
\begin{align}
\label{eq:averagechiraldisp}
\sum_{j=1}^{\D/2}\langle \Gamma \cdot\hat{m}(t)\rangle_{\overline{\Gamma_j}}
= \sum_{j=1}^{\D/2}\langle \hat{m}(t)\rangle_{\overline{\Gamma_j}}
=  \sum_{j=1}^{\D/2}\oint\frac{{\rm d}k}{2\pi}  \SP_{j}[-1+\cos(2tE_{j})]
= - \frac{\W}{2}+\sum_{j=1}^{\D/2}\oint\frac{{\rm d}k}{2\pi}  \SP_{j}\cos(2tE_{j})= - \frac{\W}{2}+\ldots.
\end{align} 
plus oscillatory terms (indicated by $\ldots$), which tend to zero in the long time limit. We also find a similar result for the mean chiral displacement $ \langle \widehat{\Gamma m}(t) \rangle $ (see App.~\ref{sec:app:meandisplacement} for details):
\beq\label{eq:meanchiraldisp}
  \sum_{j=1}^{\D/2}\langle \widehat{\Gamma m}(t) \rangle_{\overline{\Gamma_j}}
   =\sum_{j=1}^{\D/2}\langle \widehat{\Gamma m}(t) \rangle_{\overline{\psi_j}}
= \sum_{j=1}^{\D/2}\oint\frac{{\rm d}k}{2\pi}  \SP_{j}[1-\cos(2tE_{j})]
= \frac{\W}{2}+\ldots,
\eeq

The expressions \eqref{eq:averagechiraldisp} and \eqref{eq:meanchiraldisp} are invariant under the change of $j$ to $-j$, as the skew polarization is invariant under such change. Therefore, we can compute the traces  over all the $\mathcal{D}$ chiral localized states:
\beq\label{fullTraceGammaM}
- {\rm Tr}[\Gamma \cdot \hat m(t)] ={\rm Tr}[ \widehat{\Gamma m}(t)] 
= 2 \sum_{j =1}^{\D/2}\oint\frac{{\rm d}k}{2\pi}  \SP_{j}[1-\cos(2tE_{j})]
= \W+\ldots,
\eeq
As the trace does not depend of the choice of the basis, these results imply that a trace taken on {\it any} set of $\D$ vectors forming a complete basis of the unit cell will converge to the winding number $\mathcal{W}$ in the long time limit. These expressions constitute the main results of our work. \edit{Let us note that the method also works for systems with periodic boundary conditions. In the latter case however, the time of measurement should be chosen sufficiently long such that the oscillatory term becomes negligible, but at the same time sufficiently short such that the walker does not reach the lattice site where the position operator has a discontinuity.}

\section{The SSH$_4$ model}
\label{sec:SSH4}

\begin{figure}[t]
	\begin{center}
		\includegraphics[scale=1]{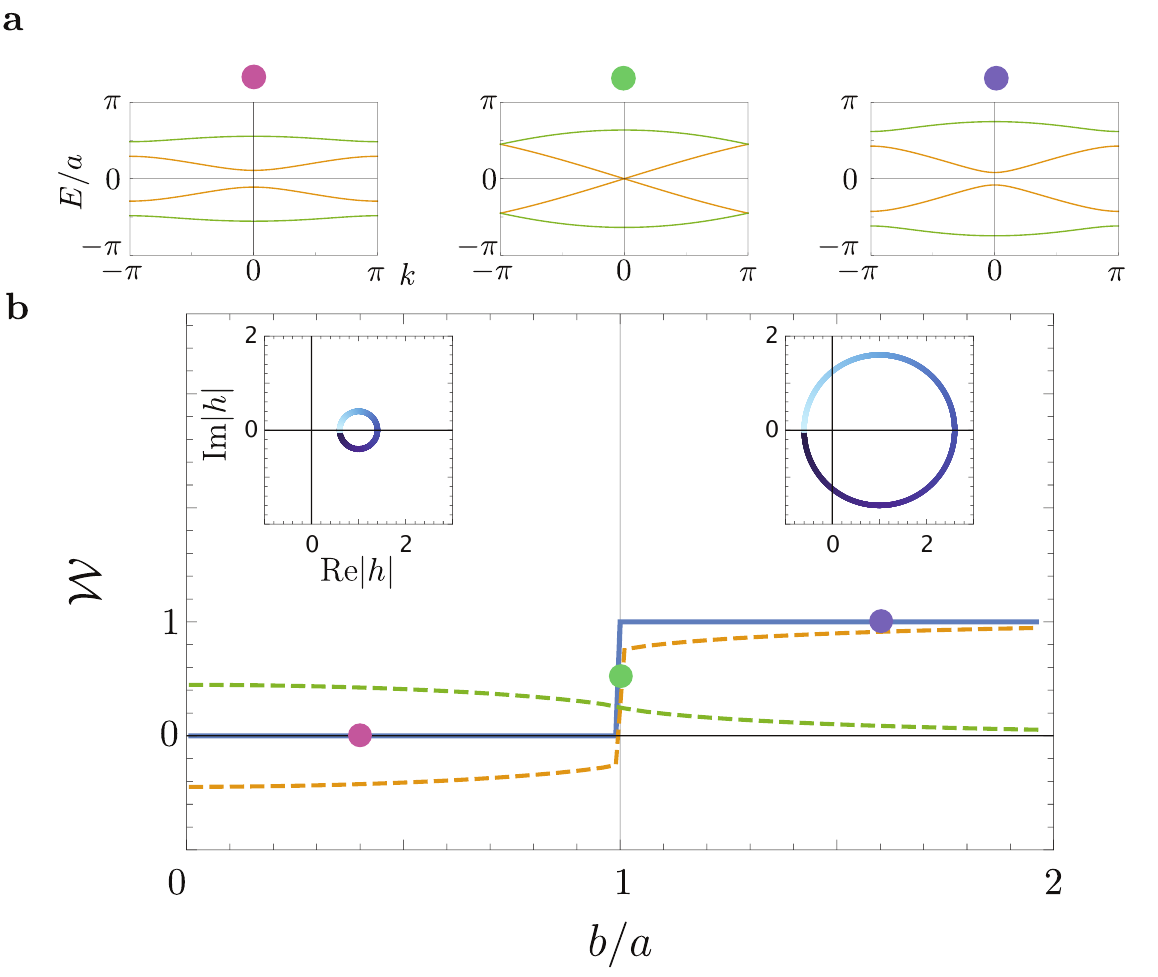} 
		\caption{\label{fig:SSH4W}
			{\bf Spectrum and topology of the SSH$_4$ model}. \textbf{a.} Dispersions for $a=c=d$, and $b/a=0.4,1,1.6$ (from left to right). The model is in the trivial phase for $b d < a c$ ($\W=0$, left), and in the non-trivial phase for $b d > a c$ ($\W=1$, right). At the critical point, the two central bands touch at $E=k=0$ (center). Eventual gap closings between the other bands (such as the one visible in the central figure, at $E/a\approx \pm\pi/2$) have no topological relevance for this model.
			\textbf{b.}  Winding number for the SSH$_4$ model with $a=c=d$, as a function of $b/a$. The yellow and green dashed lines are the separate contributions of the two pairs of chiral partners to the total winding number, respectively 
$\oint \frac{dk}{\pi}\mathcal{S}_1$ and $\oint \frac{dk}{\pi}\mathcal{S}_2$, while the blue solid line is the actual winding number, given by their sum. The insets display the value of the determinant of $h(k)$ in the complex plane, as $k$ is varied between $0$ (blue) to $2\pi$ (white).}
			\end{center}
\end{figure}

\subsection{Hamiltonian and winding number}
We now discuss an example of chiral model with $\mathcal{D}=4$, which is a direct generalization of the SSH model. The SSH$_4$ model is described by a non-interacting Hamiltonian with nearest-neighbour hoppings, as shown in Fig.~\ref{fig:sketch_models}b. The system is a Bravais lattice with a four atom unit cell of sites $A_1,B_1,A_2,B_2$. The intra-cell hoppings are $a$, $b$ and $c$, the inter-cell hopping is $d$. The Hamiltonian defines two sublattices, containing two sites each: $A=\{A_1,A_2\}$, and $B=\{B_1,B_2\}$. Since the Hamiltonian contains no term acting within a given sublattice, the model is chiral for arbitrary values of $ \{ a, b, c, d \} $. The model belongs therefore to class AIII when the tunnelings are complex numbers, while it belongs to the more constrained class BDI if all tunnelings are purely real. We will for simplicity restrict ourselves to the latter case, but note that our results hold for all 1D chiral models, i.e., also for AIII ones, such as the one considered in Ref.~\cite{Mondragon-Shem2014}. Finally, note that for $a=c$ and $d=b$ the SSH$_4$ reduces to the usual SSH model, shown in Fig.~\ref{fig:sketch_models}a. In the canonical basis  $\{\psi_{A_1},\psi_{A_2},\psi_{B_1},\psi_{B_2}\}$, the Bloch Hamiltonian assumes the off-diagonal form
\beq\label{H_SSH4_offdiag}
H(k)=\left(\begin{tabular}{cc}
	0 & $h^\dagger(k)$  \\
	$h(k)$ & 0
\end{tabular}\right)
=\left(\begin{tabular}{cccc}
	0 & 0  &  $a$ & $d e^{-ik}$  \\
	0 & 0 & $b$ & $c$ \\
	$a$ & $b$ &  0 & 0 \\
	$d e^{ik}$ & $c$ & 0 & 0
\end{tabular}\right).
\eeq
The energy spectrum and the eigenvectors of the different bands may be found analytically (see App.~\ref{sec:eigensystem_SSH4} for details). The corresponding windings are computed from Eq.~\eqref{gammaAO}, and by direct integration one finds $\W=0$ when $ac>bd$, and $\W=1$ when $ac<bd$. Figure~\ref{fig:SSH4W}a shows the energy spectrum for $a=c=d$ and for different values of $b$. The gap closing appears at $b=1$, as it is the case for the SSH model. Figure~\ref{fig:SSH4W}b shows the winding number in terms of $b$ (solid line).  The yellow and green dashed lines are the separate contributions of the two pairs of chiral partners to the total winding number. The separate conrributions are not quantized, but their sum is. Finally, the insets of the Figure show parametric plots of the determinant of $h$, which performs a circle in the complex plane as $k$ traverses the Brillouin zone. In the topological phase, the circle contains the origin (right inset) whereas in the trivial phase the circle does not contain the origin (left inset).

\subsection{Measure of the winding number in real space}

As a benchmark of the results derived in Sec.~\ref{eq:meas_winding}, we proceed by measuring the winding number through the mean chiral displacement. We consider a finite system of $200$ unit cells, we prepare localized initial states at the center of the chain $m=0$, and we let them evolve. In particular, we choose as initial states two different bases of the internal space: the chiral basis, and an arbitrary basis. At each time $t$, we compute (minus) the trace of the chiral average displacement $-{\rm Tr}[\Gamma \cdot \hat m(t)] $ of Eq.~\eqref{eq:averagechiraldisp} on the chiral basis, and the trace of the mean chiral displacement ${\rm Tr}[ \widehat{\Gamma m}(t)] $ of Eq.~\eqref{fullTraceGammaM} on an arbitrary basis. With the choice of the unit cell $\{\psi_{A_1},\psi_{B_1},\psi_{A_2},\psi_{B_2}\}$, in real space these operators are simply represented by the diagonal matrices $\hat{m} = {\rm diag}(\ldots, 1, 1, 1, 1, 2, 2, 2, 2, \ldots)$ and
$\widehat{\Gamma m} = {\rm diag}(\ldots, 1, -1, 1, -1, 2, -2, 2, -2, \ldots)$. 

\begin{figure}[t]
\begin{center}
\includegraphics[width=.45\columnwidth]{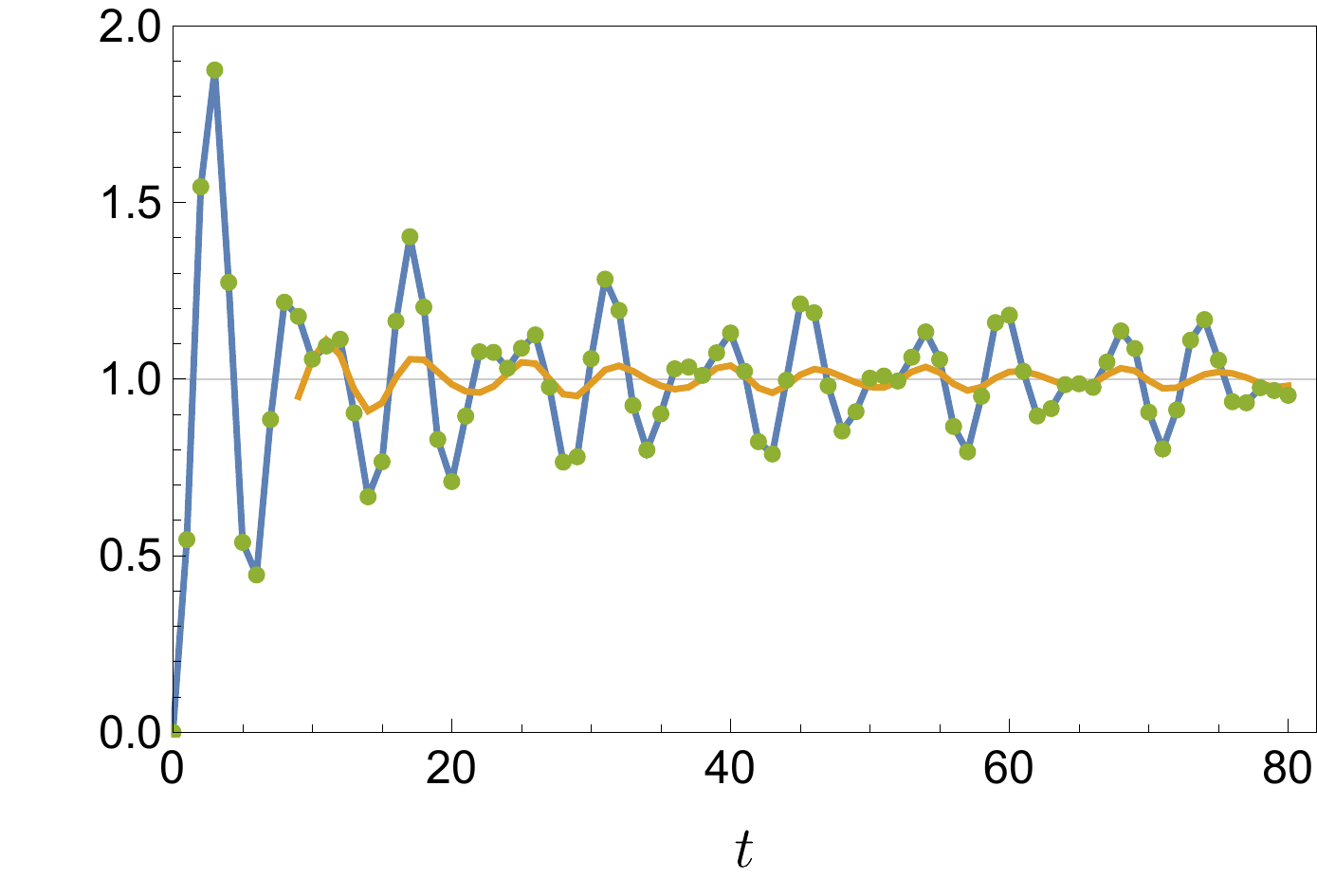} 
\caption{\label{fig:C_4}
{\bf Mean chiral and chiral average displacements of the SSH$_4$ model.} Parameters are chosen in the topological phase: $\{a,b,c,d\}=\{1, 2.5, 0.3, 0.6\}$, so that $bd>ac$ and $\W=1$. There are 2 completely superposed series of dots, showing the results of the two observables discussed in the text, and the blue line shows the analytical result, Eq.~\eqref{fullTraceGammaM}. The yellow line shows a sliding average of the data, which rapidly converges to the expected value of $1$. }
\end{center}
\end{figure}

Figure~\ref{fig:C_4} shows the results of the numerical simulations. The two traces in the different bases are superimposed (green dots), and match perfectly with the theoretical curve (blue curve) given in Eq.~\eqref{fullTraceGammaM}.  
In the figure we also show a sliding average of the data over ten points (orange curve), which shows a smoother and quicker convergence to the winding number.

Finally, let us note that the simplest procedure which yields the desired result (the winding) is to follow Eq.~\eqref{eq:averagechiraldisp} and take the sum of the mean displacement measured over two orthogonal states which are completely localized on the central unit cell, and which form a complete basis of the left sublattice (the one corresponding to the +1 eigenvalue of the chiral operator). Minus two times this quantity will give the result plotted in Fig.~\ref{fig:C_4}.
Explicitly, e.g., two states of the form $\bar{\Psi}_a=(0,\ldots,0,{\bf   1,0,0,0},   0,\ldots,0)$
and $\bar{\Psi}_b= (0,\ldots,0,  {\bf 0,1,0,0},   0,\ldots,0),$
where the four central numbers (marked in bold) indicate the amplitudes on the cell with coordinate $m=0$ in the basis where the chiral operator is (1,1,-1,-1).

\section{Driven SSH$_4$ model}
\label{sec:Floquet_SSH_4}

\begin{figure}[t]
\begin{center}  
\includegraphics[scale=1]{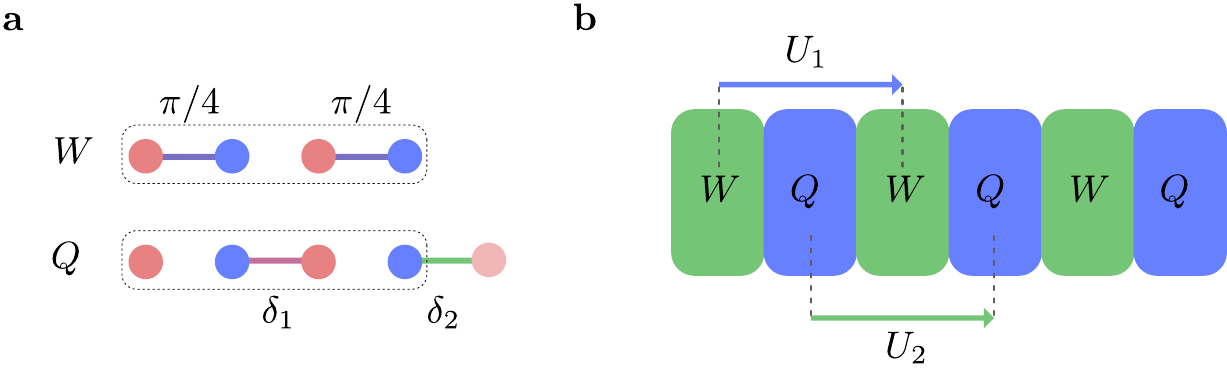} 
\caption{\label{fig:QW} 
{\bf The driven SSH$_4$ model.}
\textbf{a.} Scheme of the unit step of the driven SSH$_4$ model, or quantum-walk with four dimensional coin. The four sites of each unit cell correspond to the coin states $\ket{A_1},\ket{A_2}$ (red spots), and $ \ket{B_1}, \ket{B_2}$(blue spots).  The dynamics is obtained by the repeated application of two unitary operators: $W$ is a rotation acting on the odd sites of the lattice (which are all intracell), while $Q$ acts on the even sites, thereby coupling different cells. \textbf{b.} Scheme of the quantum-walk protocol.
In a periodically driven system, the choice of the initial instant of the time period, i.e. the time-frame, is arbitrary. The single-step unitary operators $U_1$ and $U_2$ correspond to two inversion-symmetric time-frames.}
\end{center}
\end{figure}

In this section, we show that our detection method works even for Floquet systems. In order to do so, we consider a periodically-driven version of the SSH$_4$ model, where even and odd tunnelings are turned on and off in a periodic sequence.
More specifically, a single period of the evolution is generated by the one-step operator $U$ given by a composition of the two unitary operators $W$ and $Q$, as shown in Fig.~\ref{fig:QW}.
The operator $W$ acts on odd links (see Fig.~\ref{fig:QW}a), therefore only within each unit cell:
\begin{equation}
W= e^{-i \frac{\pi}{4} S_{0x} }
\end{equation}
where $S_{ij}=\sigma_i \otimes \sigma_j$. 
On the other hand, the operator $Q$ acts on even links, therefore both within a given cell, and between two consecutive cells:
\begin{align}\label{translation}
Q \cdot (\ket{A_1} \otimes \ket{m})&=\cos\frac{\delta_2}{2}\ket{A_1} \otimes \ket{m}- i \sin\frac{\delta_2}{2} \ket{B_2} \otimes \ket{m-1} \\ \nonumber
  Q \cdot (\ket{B_1} \otimes \ket{m})&=\left(\cos\frac{\delta_1}{2}\ket{B_1}- i \sin\frac{\delta_1}{2} \ket{A_2} \right) \otimes \ket{m} \\ \nonumber
Q \cdot (\ket{A_2} \otimes \ket{m})&=\left(\cos\frac{\delta_1}{2}\ket{A_2}- i \sin\frac{\delta_1}{2} \ket{B_1}\right) \otimes \ket{m}\\ \nonumber
 Q \cdot (\ket{B_2} \otimes \ket{m})&=\cos\frac{\delta_2}{2}\ket{B_2} \otimes \ket{m}- i \sin\frac{\delta_2}{2} \ket{A_1} \otimes \ket{m+1}.
 \end{align}
The scheme proposed above effectively realizes a discrete-time quantum walk with a four-dimensional coin, a generalization of the usual topological quantum walk with two-dimensional coin \cite{Kitagawa2012, Asboth2013, Barkhofen2016, Cardano2016, Cardano2017, Flurin2017}.

In order to completely characterize the topology of this driven model, we follow the method proposed in Ref.~\cite{Asboth2013}, and recently implemented in Ref.~\cite{Cardano2017}. Different choices of timeframes, i.e., of the initial instant of the periodic cycle, yield effective Hamiltonians $H^{\rm (eff)}=i\log U$ with the same set of eigenvalues, but different eigenvectors, and therefore possibly different windings. Here we consider the two chiral inversion-symmetric timeframes defined by the evolution operators $U_1=\sqrt{W} Q \sqrt{W}$ and $U_2=\sqrt{Q} W  \sqrt{Q}$, shown schematically in Fig.\ \ref{fig:QW}b.

\begin{figure}
\begin{center}   
\includegraphics{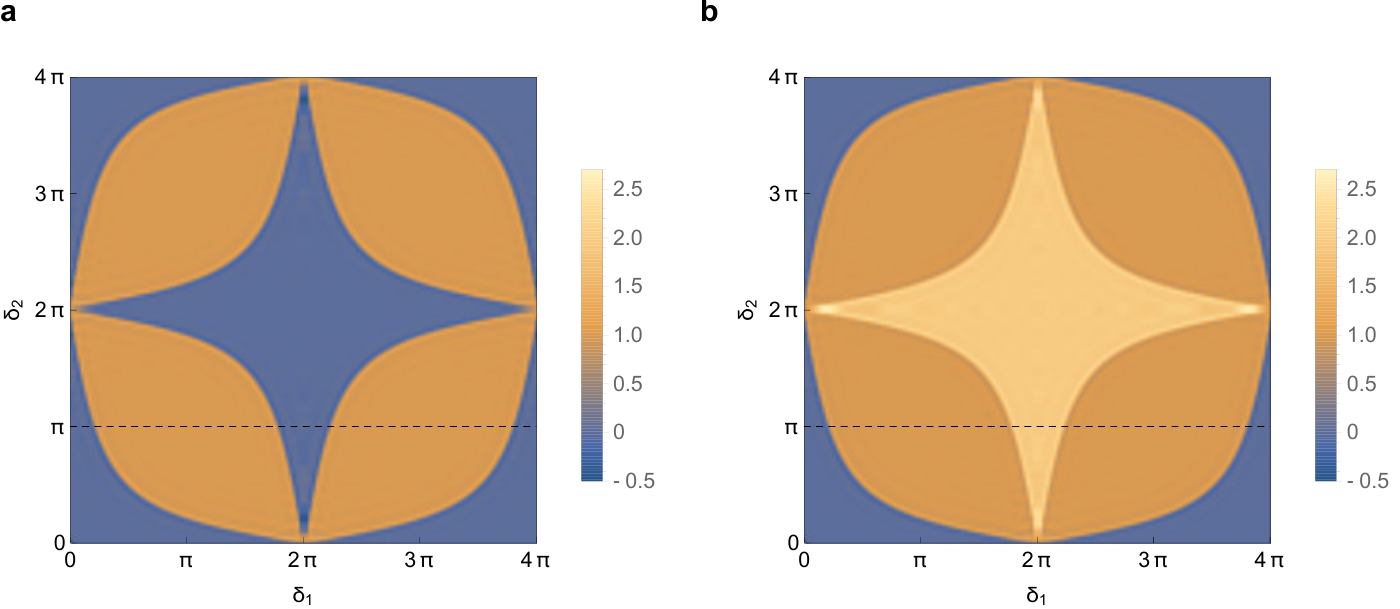} 
\caption{\label{fig:tracefloquet1}
{\bf Mean chiral displacement of the driven SSH$_4$ model.} Temporal average of the MCD, obtained performing a sliding average between the 30$^{\rm th}$ and the 50$^{\rm th}$ step of the walker's evolution, as a function of the parameters $\delta_1$ and $\delta_2$ characterizing the action of the Q-plate, for the time-frames $U_1$ ({\bf a}) and $U_2$ ({\bf b}).}
\end{center}
\end{figure}

\begin{figure}[h]
\begin{center}  
\includegraphics{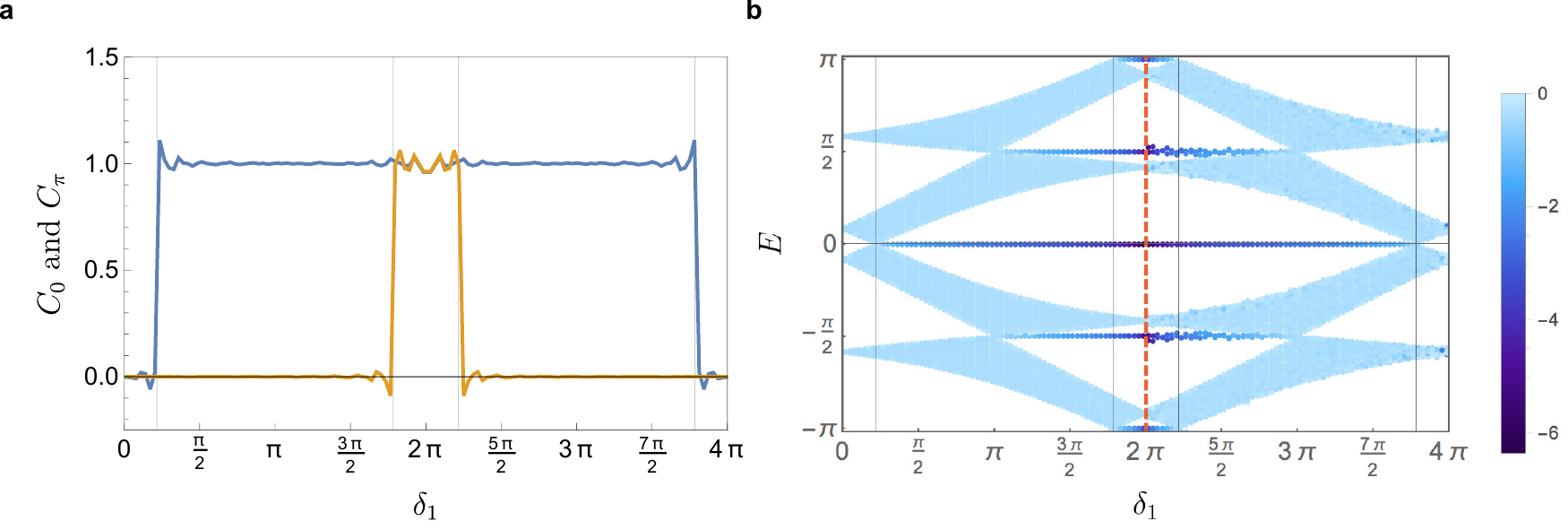}  
\caption{\label{fig:effen}
  {\bf Bulk-edge correspondence for the driven SSH$_4$.} 
  {\bf a.} invariants $C_0=(\W_1 +\W_2)/2$ (blue) and $C_\pi=(\W_1 -\W_2)/2$ (orange) for $\delta_2 =\pi$. 
Continuous lines are obtained from the traces of the mean chiral displacements measured in timeframes $U_1$ and $U_2$ shown in Fig.~\ref{fig:tracefloquet1}.
{\bf b.} spectrum with open-boundary conditions  on a lattice containing $(2L+1)=21$ effective cells, varying $\delta_1$ at fixed $\delta_2 =\pi$. 
The color coding of the spectrum indicates the degree of localization $\log_{10}(1-\vert \langle m\rangle \vert/L)$ of each eigenstate; light (dark) colors indicate bulk (edge) states. 
For $\delta_1>2\pi$, we have added weak chiral-preserving disorder (see text for details)  with $\Delta=0.6$, showing explicitly that the edge states with $E=\pm\pi/2$ are localized, but not topologically protected.
Comparing the left and right image, it is easy to see that $C_0$ and $C_\pi$ predict respectively the number of edge states with 0- and $\pi-$energies. }
\end{center}
\end{figure}

The time evolution of a wavepacket after $t$-timesteps of protocol $j$ ($j=\{1,2\}$) is as usual governed by the evolution operator $(U_j)^t$. 
Sliding averages over the long-time behavior of the trace of the chiral displacement are shown in Fig.~\ref{fig:tracefloquet1}.  
We obtain the complete topological characterization of the system in terms of the invariants $C_0\equiv(\W_1 +\W_2)/2 $ and $C_\pi \equiv(\W_1 -\W_2)/2$, where $\W_j$ is the winding of protocol $U_j$. Figure~\ref{fig:effen}a shows $C_0$ and $C_\pi$ on the line $\delta_2=\pi$. 
To illustrate the ``bulk-boundary" correspondence for this model,
in Fig.~\ref{fig:effen}b we show the energy spectrum and the degree of localization of eigenstates in a chain with open boundary conditions. 
Comparing the two panels, it may be seen that the invariants $C_0$ and $C_\pi$ converge, respectively, to the number of edge states with energy equal to 0 and to $\pi$, which are the ones protected by chiral symmetry in a driven system. 
Finally, Fig.\ \ref{fig:effen}b also shows the presence of edge states with energy $\pi/2$.
These states are not protected by the chiral symmetry, and therefore not robust against (chiral-preserving) disorder. In order to illustrate this fact, we add a spatial disorder in the operator $W$: the hoppings of the Hamiltonian of $W$ are multiplied by a factor $(1+\epsilon)$, where $\epsilon$ is  a random number in the range$[-\Delta/2,\Delta/2]$. The right side of the energy spectrum (after the dashed line) in Fig.~\ref{fig:effen}b shows clearly that, whereas the $0$ and the $\pi-$energy states remain unaffected, the unprotected states change of energy when disorder is applied.

\edit{The effect of spatial disorder and noise on the mean chiral displacement for systems with $\D=2$ were already discussed at length in our previous publication, Ref.~\cite{Cardano2017}. In particular, there we confirmed that this observable is a robust topological marker by showing that, in presence of chiral-preserving static spatial disorder of amplitude small compared to the gap, the ensemble average of the mean chiral displacement smoothly converges to the value obtained for a clean system. In the literature it has been discussed how a different observable may be used to detect the winding of chiral non-Hermitian models~\cite{Rudner2009,Zeuner2015,Rakovszky2017}. Let us note that there are important differences between the latter proposals and ours. First, the scheme proposed in Refs.~\cite{Rudner2009,Zeuner2015,Rakovszky2017} requires the initial state to be polarized along a well-defined direction, while our method works independently of the polarization of the initial condition. Second, that method requires sublattice-dependent losses that may not be easily introduced in an experiment with ultracold atoms, or in an optomechanical system and do not work without losses. Finally, our method works even in presence of uniform losses while the method for non-Hermitian systems does not. 

At a qualitative level, systems with internal dimension $\mathcal{D}>2$ behave in no manner differently from systems with $\D=2$ in presence of disorder. Readers interested in this topic are therefore referred to our earlier publication Ref.~\cite{Cardano2017}, and to its Supplemental Information, where the matter is discussed in great detail.
}

\begin{figure}
\begin{center}
\includegraphics[scale=1]{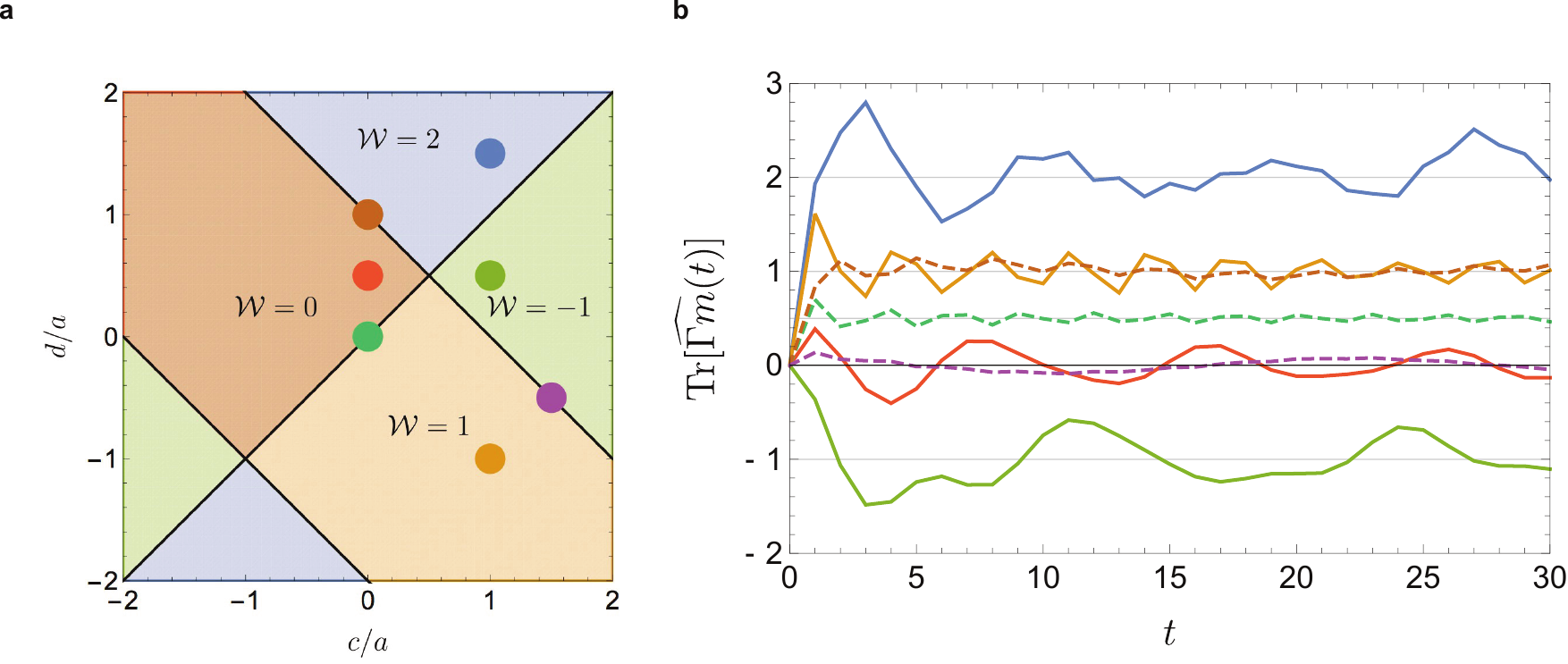}  
\caption{\label{fig:LR_SSH}
{\bf Topology of the LR-SSH model.} {\bf a.} windings, with $a=b$. {\bf b.} mean chiral displacement computed for the values of $(c,d)$ indicated by the corresponding dot in the left figure. The three dashed lines correspond to values of the parameters at the border between two phases, where the model is critical; in these cases, the observable remarkably converges to the average of the corresponding invariants.}
\end{center}
\end{figure}

\section{SSH model with staggered long range hoppings}
\label{sec:LR-SSH}
In this Section we wish to test the validity of our detection method using a system which exhibits a richer phase diagram ($\W=0,\pm 1,2$). To this aim, we study a one-dimensional chiral Hamiltonian, which is a standard SSH model with staggered nearest-neighbor hoppings $a$ and $b$, and with additional staggered third-nearest-neighbor hoppings $c$ and $d$ (that is to say, there is a hopping $c$ between sites $1$ and $4$, $d$ between sites $2$ and $5$, $c$ between $3$ and $6$, and so on). The model is shown schematically in Fig.~\ref{fig:sketch_models}c, and given its long-range character we refer to it as the LR-SSH model. 

The model can be written in momentum space when using a two-atom unit cell. The corresponding Hamiltonian is a 2*2 matrix,
\beq
H_{\rm LR}=\left(\begin{tabular}{cc}
	0  &  $a+be^{-ik}+ce^{ik}+de^{-2ik}$\\
	$a+be^{ik}+ce^{-ik}+de^{2ik}$ & 0
\end{tabular}\right).
\eeq
 
 The winding $\W$ of this model may be computed as shown in Sec.~\ref{subsec:winding}, and it equals +2, +1, 0, or -1.
The topological phase diagram with $a=b$ is shown in Fig.~\ref{fig:LR_SSH}a. 

As shown in Fig.~\ref{fig:LR_SSH}b, the long time limit of the trace of the mean chiral displacement detects correctly the winding in all topologically distinct regions. In particular, when $c=d=0$, the model is at the critical point between the phases with $\W=0$ and $\W=1$, recovering the expected SSH result, which is critical when $a=b$. In this point, as in all other phase transitions, the mean chiral displacement converges to the intermediate (integer, or half integer) value between the windings of the neighboring phases (see dashed lines in Fig.\ \ref{fig:LR_SSH}b), as discussed for example in Ref.~\cite{Altland2015}.

\section{Experimental implementation, outlook, and conclusions}
\label{sec:outlook}

\begin{figure}
	\begin{center}
		\includegraphics[scale=1]{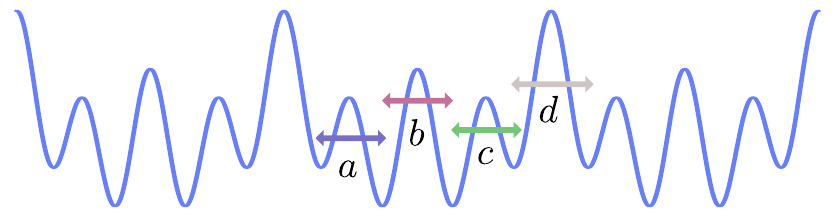} 
		\caption{\label{fig:SSH4_superlattice}
			Implementation of the SSH$_4$ model, obtained superposing optical lattices with lattice spacings  $\lambda/2$, $\lambda$, and $2\lambda$; in this configuration $a=c$, but $b$ and $d$ may be tuned independently.}
	\end{center}
\end{figure}

Various possible experimental scenarios may be envisaged to study chiral models with large internal dimensions.
For example, a $\D=4$ chiral model with ultracold atoms may be implemented by means of a suitable superlattice as it has been proposed also in Ref.\ \cite{guo2015}. Three superposed optical lattices with lattice spacings $\lambda/2$, $\lambda$, and $2\lambda$ effectively realize an SSH$_4$ model with two equal tunnelings, as shown in Fig.~\ref{fig:SSH4_superlattice}. The three lattices may be obtained from a single laser working at $\lambda_{\rm laser}=1064$nm, which once retroreflected produces an optical potential with lattice spacing $\lambda=\lambda_{\rm laser}/2$. The $\lambda/2$ lattice may be obtained by retroreflecting the frequency-doubled laser, while the one at $2\lambda$ may be obtained by crossing two $\lambda_{\rm laser}$ beams at a small angle. Otherwise, the superlattice may be by directly imprinted with a spatial light modulator (SLM) or with a digital mirror device (DMD). Driven models may be realized by periodically pulsed Hamiltonians, such as the one discussed, e.g.,  in Ref.~\cite{Rudner2013}.

Two-photon Bragg processes were used in recent experiments by the group of B.~Gadway to realize a static SSH model in momentum space \cite{Meier2016}. The independent control of each hopping allows for the possibility to engineer the long-ranged tunnelings we discussed in Sec.~\ref{sec:LR-SSH}, and driven models may be obtained by periodically modulating the amplitude of the Bragg lasers. This architecture allows also for a detailed study of the interplay between topology and disorder \cite{an2017}.

In a photonic setting, we envisage using a lattice of evanescently coupled optical waveguides, where the different hopping amplitudes correspond to different distances between the waveguides \cite{Zeuner2015, Maczewsky2017, Broome2013, Crespi2013}. The driven SSH$_4$ discussed in Sec.\ \ref{sec:Floquet_SSH_4} may be implemented by periodical modulation of the separation between the waveguides along the propagation direction \cite{Maczewsky2017,Crespi2013}, and long-ranged tunnelings may be obtained by letting the waveguides propagate out of the plane, as possible in 3D-photonic chips \cite{Crespi2016, Harris2017}.
Finally, the SSH$_4$ model may be implemented in exciton-polariton experiments, by a slight modification of the approach used by the group of A.~Amo in Ref.~\cite{StJean2017}.

Summarizing, in this work  we have generalized the notion of mean chiral displacement to chiral systems with any internal dimension $\D$, showing that when $\D>2$ the winding number is encoded in the long-time limit of the trace of the mean chiral displacement over a localized basis of the internal space. 
We analyzed three chiral models having internal dimension $\D=4$, i.e.: (i) the simplest extension of the SSH model (the SSH$_4$), (ii) its driven analogue, and (iii) the SSH model with staggered long range hoppings.  
We applied our detection method to these systems showing that it works correctly for both static and periodically-driven Hamiltonians, and that it is able to capture values of the winding which exceed 1, such as the LR-SSH. Furthermore, the detection is rapid, as the amplitude of the oscillations displayed by the mean chiral displacement is generally smaller than the length of a unit cell, and therefore only very few tunneling times are needed to identify the winding of the model. Moreover, the detection method we propose does not require filled bands, but is based on the dynamics of wavepackets, or even single particles, and therefore it is specially adapted to systems where band-filling may be problematic, like bosonic ensembles of atoms, or photonic systems. Finally, the method only requires to follow the free evolution of the system, without the need of extra resources, such as external forces, losses, or interferometric schemes. The main ingredient for the characterization of the winding is the single site detection, which can be realized with state-of-the-art experimental techniques.

Future interesting directions opened by this work include a study of the robustness of our observable against static and temporal disorder \cite{Mondragon-Shem2014}, the extension of this approach to interacting systems \cite{Gurarie2011,Manmana2012}, and eventually the extension of this proposal to other topological classes.

\acknowledgments
We thank Alessio D'Errico, Victor Gurarie, Lorenzo Marrucci, Pierre Delplace and Leticia Tarruell for enlightening discussions, and in particular J\'anos Asb\'oth for the insightful comments that motivated this work.

This work has been supported by 
Spanish MINECO (Severo Ochoa SEV-2015-0522 and FisicaTeAMO FIS2016-79508-P),
the Generalitat de Catalunya (SGR 874 and CERCA),
Fundaci\'o Privada Cellex, 
and EU grants 
OSYRIS (ERC-2013-AdG Grant 339106), 
QUIC (H2020-FETProAct-2014 641122), 
and  SIQS (FP7-ICT-2011-9 600645).
MM acknowledges funding from EU grant OSYRIS.
AD is financed by a Cellex-ICFO-MPQ fellowship.
FC acknowledges funding from Adv.~ERC grant PHOSPhOR (694683).
PM acknowledges funding from the ``Ram\'on y Cajal" program and the Simons Foundation, and the kind hospitality of the Aspen Center for Physics, where part of this work was realized, and which is supported by National Science Foundation grant PHY-1607611.


\appendix
\section{Proofs}

\subsection{Mean displacement}\label{sec:app:meandisplacement}

In terms of projectors on energy eigenstates, the mean displacement of a generic localized state $\overline{\ket{\Psi}}$ reads:
\beq
\langle \hat{m}(t)\rangle_{\overline{\Psi}}
=\oint\frac{{\rm d}k}{2\pi}\sum_{j,j'=\pm 1,...,\pm\D/2}\bra{\Psi}P_{j}U^{-t}(i\partial_k)U^tP_{j'}\ket{\Psi}.
\eeq
Using $\partial_k U^t P_{j'}=e^{-itE_{j'}}[(-i t \partial_k E_{j'})P_{j'} + \ket{\psi_{j'}'}\bra{\psi_{j'}} + \ket{\psi_{j'}}\bra{\psi_{j'}'}]$, we have
\begin{align}\label{projectors_meanDisp}
P_{j} U^{-t}\partial_k U^tP_{j'}&=e^{itE_{j}}P_{j}\partial_k U^t P_{j'}\\\nonumber
&=\delta_{jj'} [-it(\partial_k E_j)P_{j}+\ket{\psi_j}\bra{\psi_{j}'}]+e^{it(E_{j}-E_{j'})}P_{j}\ket{\psi_{j'}'}\bra{\psi_{j'}}.
\end{align}

\subsubsection{Mean displacement in $\D=2$}

For $\D=2$, we will denote with $+/-$ the positive/negative energy eigenstates, so that the skew polarization is simply $\SP=i\bra{\psi_+}\psi_-'\rangle$.
Multiplying Eq.\ \eqref{projectors_meanDisp} by $(+i)$, using 
$\ket{\psi_+}\bra{\psi_{-}}-\ket{\psi_-}\bra{\psi_{+}} = Q \Gamma$, and inserting a completeness relation 
$P_++P_-=\mathbb{I}$ after $\ket{\psi_j}\bra{\psi_{j}'}$, one finds:
\begin{align}\label{meanDispTwoDimProjectors}
\langle \hat{m}(t)\rangle_{\overline{\Psi}}
=\oint\frac{{\rm d}k}{2\pi}\Big\{
t\partial_kE_+\mean{Q}_\Psi 
+\sin(2tE_+)\SP \mean{i Q\Gamma}_\Psi
- \SP[1-\cos(2tE_+)]\mean{\Gamma}_\Psi 
\Big\}.
\end{align}
Now we use the following relations:
\begin{itemize}
\item 
 $Q={\bf n}\cdot\vectgr{\sigma}$
\item $i Q\Gamma =-n_y \sigma_x + n_x \sigma_y$
\item $ \SP\mean{i Q\Gamma}=\mean{\partial_k {\bf n}\cdot\vectgr\sigma}/2$
\item
 $\SP[1-\cos(2tE_+)]=2 \SP\sin^2(tE_+)=\sin^2(tE_+)({\bf n}\times\partial_k{\bf n})$.
\end{itemize} 
Then Eq.\ \eqref{meanDispTwoDimProjectors} may be written as:
\begin{align}
\langle \hat{m}(t)\rangle_{\overline{\Psi}}
=\oint\frac{{\rm d}k}{2\pi}\Big\{
t\partial_kE_+\mean{{\bf n}\cdot \vectgr\sigma}_\Psi 
+\frac{\sin(2tE_+)}{2}\mean{\partial_k{\bf n}\cdot\vectgr\sigma}_\Psi
- \sin^2(tE_+)({\bf n}\times\partial_k{\bf n})\mean{\Gamma}_\Psi 
\Big\}.
\end{align}
This result coincides with the one given in the Supplemental Material of  Ref.\ \cite{Cardano2017}.
 
In the particular case of a chiral localized state $\overline{\ket{\Gamma_{j}}}$, only the last term of Eq.\ \eqref{meanDispTwoDimProjectors} survives. This comes from the facts that $\mean{Q}_{\Gamma_j}=0$ and $\mean{Q\Gamma}_{\Gamma_j}=0$.
 
On the other hand, for a localized state built as flat superposition of states in a single band $\overline{\ket{\psi_{j}}}$, Eq.\ \eqref{meanDispTwoDimProjectors} gives 0. This comes from the facts that $\oint dk \partial_k E_+ \mean{Q}_{\psi_j}= sign(j)\oint dk \partial_k E_+ =0$, $\mean{Q\Gamma}_{\psi_j}=0$ 
 and $\mean{\Gamma}_{\psi_j}=0$. 
\subsubsection{Mean displacement in $\D>2$}\label{subsubsec:meandisp4}

For $\D>2$, the mean displacement at time $t$, starting from a generic  localized state $\overline{\ket{\Psi}}$ reads:
\begin{align}\label{meanDispOverlinePsi}
\langle \hat{m}(t)\rangle_{\overline{\Psi}} 
&=\sum_{j=1}^{\D/2}
\oint\frac{{\rm d}k}{2\pi} \Big\{
t\partial_kE_{j}\mean{Q_{j}}_\Psi 
+\SP_{j}\sin(2tE_{j}) \mean{i Q_{j}\Gamma_{j}}_\Psi
- \SP_{j}[1-\cos(2tE_{j})]\mean{\Gamma_{j}}_\Psi 
\Big\} \\\nonumber
&+\sum_{j, j'=\pm 1,\ldots,\D/2 \textrm{ and }|j|\neq|j'|}
\oint\frac{{\rm d}k}{2\pi} i\bra{\psi_j}\psi_{j'}'\rangle \langle \Psi\ket{\psi_j}\bra{\psi_{j'}}\Psi\rangle e^{i t (E_j-E_{j'})}.
 \end{align}
It can be shown that the terms arising from the second summation give rise to a purely real number, in agreement with the fact that the result is the expectation value of a Hermitian operator.

Noting that $Q_j\Gamma_j=\ket{\psi_j}\bra{\psi_{-j}}-\ket{\psi_{-j}}\bra{\psi_{j}}$, it is easy to see that the states $\overline{\ket{\psi_j}}$ are again stationary, as expected.
On the other hand, for a chiral localized state $\overline{\ket{\Gamma_j}}$, Eq.~\eqref{meanDispOverlinePsi} gives:
\begin{align}
\langle \hat{m}(t)\rangle_{\overline{\Gamma_j}}
 &=-{\rm sgn}(j)\oint\frac{{\rm d}k}{2\pi}\SP_{j}[1-\cos(2tE_{j})],
\end{align}
which proves Eq.\ \ref{eq:averagechiraldisp}:
\begin{align}
\sum_{j=1}^{\D/2}\langle \hat{m}(t)\rangle_{\overline{\Gamma_j}}
= \sum_{j=1}^{\D/2}\langle \Gamma \cdot\hat{m}(t)\rangle_{\overline{\Gamma_j}}
= - \sum_{j=1}^{\D/2}\oint\frac{{\rm d}k}{2\pi}  \SP_{j}[1-\cos(2tE_{j})],
\end{align} 
 
\subsection{Mean chiral displacement}
\label{sec:app:meanchiraldisplacement}
In terms of projectors on energy eigenstates, the mean chiral displacement of a generic localized state $\overline{\ket{\Psi}}$ reads:
\beq
\langle \widehat{\Gamma m}(t)\rangle_{\overline{\Psi}}
=\oint\frac{{\rm d}k}{2\pi}\bra{\Psi} U^{-t}\Gamma (i\partial_k)U^t\ket{\Psi}.
\eeq
We have
\begin{align}\label{ChiralProj}
P_{j}[U^{-t}\Gamma \partial_k U^t]P_{j'} &=\delta_{jj'}\left[P_{j}\Gamma\partial_k \frac{e^{i2tE_j}}{2}+e^{i2tE_j}\ket{\psi_{j}}\bra{\psi_{-j}'}\right]
+e^{it(E_j-E_{j'})}\ket{\psi_j}\bra{\psi_{-j}}\psi_{j'}'\rangle\bra{\psi_{j'}}=\\\nonumber
&=\delta_{jj'}\left[P_{j}\Gamma\partial_k \frac{e^{i2tE_j}}{2}+e^{i2tE_j}\ket{\psi_{j}}\bra{\psi_{-j}'}\right]
-e^{it(E_j-E_{j'})}\ket{\psi_j}\bra{\psi_{-j}'}P_{j'}.
\end{align}

\subsubsection{Mean chiral displacement in $\D=2$}
For $\D=2$, the mean chiral displacement at time $t$, starting from a generic  localized state $\overline{\ket{\Psi}}$ reads: 
\begin{align}
\langle \widehat{\Gamma m}(t)\rangle_{\overline{\Psi}}
&=\oint\frac{{\rm d}k}{2\pi}\Big\{
\SP[1-\cos(2tE_+)]\left\langle\mathbb{I}\right\rangle_\Psi+\frac{1}{2}\partial_k\left[\left\langle\Gamma\right\rangle_\Psi \cos(2tE_+)+\left\langle iQ\Gamma\right\rangle_\Psi\sin(2tE_+)\right]
\Big\}=\\\nonumber
&=\oint\frac{{\rm d}k}{2\pi}\SP[1-\cos(2tE_+)]=\oint\frac{{\rm d}k}{2\pi}\SP\frac{\sin^2(tE_+)}{2}=\oint\frac{{\rm d}k}{2\pi}\sin^2(tE_+)({\bf n}\times\partial_k{\bf n}).
\end{align}
This expression coincides with the one given in Ref.\ \cite{Cardano2017}.

\subsubsection{Mean chiral displacement in $\D>2$}

Let us now define the projector on the subspace of chiral-partner eigenstates,
\beq
R_j=P_j+P_{-j},\qquad  \textrm{so that }\sum_{j=1}^{\D/2}R_j=\mathbb{I}.
\eeq
When $\D>2$, we find that Eq.\ \eqref{ChiralProj} multiplied by $i$ gives the sum of two terms, a term $A$ which acts in the subspace of chiral partner states ($|j|=|j'|$) and a term $B$ which acts in the subspace of the states with $|j| \neq|j'|$.
\begin{align}
A&=\sum_{j=1}^{\D/2} \SP_j [1-\cos(2 t E_j)] R_j + i R_j\Gamma\partial_k\left[ \frac{\cos(2tE_j)}{2}\right]
- Q_j \Gamma\partial_k\left[\frac{\sin(2tE_j)}{2}\right]-i Q_j\SP_j \sin(2tE_j)\\\nonumber
&=\sum_{j=1}^{\D/2} \SP_j [1-\cos(2 t E_j)] R_j + \partial_k\left[ i\Gamma_j \frac{\cos(2tE_j)}{2}
- Q_j\Gamma \frac{\sin(2tE_j)}{2}\right],
\end{align}
where we have used the facts that $R_j\Gamma=\Gamma_j$, $\partial_k \Gamma=0$ and $i  Q_j S_j = \partial_k (Q_j \Gamma) /2$.
And
\begin{align}
B= \sum_{j, j'=\pm 1,\ldots,\D/2 \textrm{ and }|j|\neq|j'|} 
i\bra{\psi_{-j}}\psi'_{j'}\rangle \ket{\psi_j}\bra{\psi_{j'}}e^{i t (E_j-E_{j'})}.
\end{align}
The term $B$ has no diagonal term between chiral partners, and is purely oscillatory, so for generic $E_j$ and $E_{j'}$
 it will average to zero in the long time limit.
Once integrated over the whole Brillouin zone the total derivative contained in A vanishes, so that the final result is
\begin{align}
\langle \widehat{\Gamma m}(t)\rangle_{\overline{\Psi}}
=\oint\frac{{\rm d}k}{2\pi}\left\langle B+\sum_{j=1}^{\D/2} \SP_j [1-\cos(2 t E_j)] R_j\right\rangle_\Psi.
\end{align}

For the states $\overline{\ket{\psi_j}}$ and $\overline{\ket{\Gamma_j}}$, we have that $\langle B\rangle_{\psi_j}=\langle B\rangle_{\Gamma_j}=0$ and $\langle R_j\rangle_{\psi_j'}=\langle R_j\rangle_{\Gamma_j'}=\delta_{jj'}$. This proves Eq.\ \eqref{eq:meanchiraldisp}: 
\beq\label{sumMCDoverPositiveEnergies}
\sum_{j=1}^{\D/2} \langle \widehat{\Gamma m}(t)\rangle_{\overline{\Gamma_j}}=\sum_{j=1}^{\D/2} \langle \widehat{\Gamma m}(t)\rangle_{\overline{\Psi_j}}=\oint\frac{{\rm d}k}{2\pi} \sum_{j=1}^{\D/2} \SP_j [1-\cos(2 t E_j)].
\eeq

The mean chiral displacement of a generic localized state, with support on all bands, in the long time-limit would instead be given by:
\beq
\lim_{t\rightarrow\infty} \langle \widehat{\Gamma m}(t)\rangle_{\overline{\Psi}}
= \oint\frac{{\rm d}k}{2\pi} \sum_{j=1}^{\D/2} \SP_j \langle R_j\rangle_{\Psi},
\eeq
which, differently from the case $\D=2$, is not a multiple of the winding number. 
\section{Eigensystem of the SSH$_4$ model}\label{sec:eigensystem_SSH4}

Given a generic block anti-diagonal matrix 
$M=\left(\begin{tabular}{cc}
0 & $M_{12}$  \\
$M_{21}$ & 0
\end{tabular}\right)$, we have $M^2=\left(\begin{tabular}{cc}
$M_{12}M_{21}$ & 0   \\
0 & $M_{21}M_{12}$
\end{tabular}\right)$. The eigenvalues of $M$ therefore are the square roots of the eigenvalues of $\hat{M}=M_{12}M_{21}$.
Thus, if we start from the SSH$_4$ Hamiltonian written in its completely off-diagonal form (in the canonical chiral eigenbasis), we have $H^2=\left(\begin{tabular}{ll}
$\hat{h}$ & 0  \\
0 & $\tilde{h}$
\end{tabular}\right)$, with
\beq
\hat{h}=h^\dagger .h=\left(\begin{tabular}{cc}
	$a^2+d^2$ & $ab+cde^{-ik}$  \\
	$ab+cde^{ik}$ & $b^2+c^2$
\end{tabular}\right),
\eeq
and $\tilde{h}=h.h ^\dagger$.
If we denote by $\lambda_{1}^2$ and $\lambda_{2}^2$ the two eigenvalues of $\hat{h}$, the eigenvalues of the Hamiltonian are simply given by their square roots: 
\beq
\lambda_{\pm1}=\pm \lambda_1=\pm\sqrt{\frac{T}{2}-\sqrt{\frac{T^2}{4}-\hat{D}}},\qquad
\lambda_{\pm2}=\pm \lambda_2=\pm\sqrt{\frac{T}{2}+\sqrt{\frac{T^2}{4}-\hat{D}}},
\eeq
where $T=a^2+b^2+c^2+d^2$ and $\hat{D}=a^2c^2+b^2d^2-2abcd\cos(k)$ are respectively the trace and determinant of $\hat{h}$,
and $|\lambda_{\pm1}|<|\lambda_{\pm2}|$. The topological phase transition of the SSH$_4$ model takes place when $ac=bd$ and $k=0$, where $\lambda_{\pm1}=0$.

In order to find the eigenvectors of $H$, let us first consider the eigenvectors of $H^2$. 
Provided that $e^{ik}\neq - ab/cd$, we have $H^2\ket{\hat{h}_l}=\lambda_l^2\ket{\hat{h}_l}$ (for $l=1,2$) with:
\beq
\ket{\hat{h}_1}=\frac{1}{\sqrt{\langle\hat{h}_1\ket{\hat{h}_1}}}
\left(\begin{tabular}{c} 
	$\lambda_1^2-(b^2+c^2)$ \\ $ab+cd e^{ik}$ \\0 \\ 0
\end{tabular}\right),\qquad
\ket{\hat{h}_2}=\frac{1}{\sqrt{\langle\hat{h}_2\ket{\hat{h}_2}}}
\left(\begin{tabular}{c} 
	$\lambda_2^2-(b^2+c^2)$ \\ $ab+cd e^{ik}$ \\ 0 \\ 0 
\end{tabular}\right).
\eeq
Similarly, provided that $e^{ik}\neq - bc/ad$, we have $H^2\ket{\tilde{h}_l}=\lambda_l^2\ket{\tilde{h}_l}$, with:
\beq
\ket{\tilde{h}_1}=\frac{1}{\sqrt{\langle\tilde{h}_1\ket{\tilde{h}_1}}}
\left(\begin{tabular}{c} 
	0\\0\\$\lambda_1^2-(c^2+d^2)$ \\ $bc+ad e^{ik}$
\end{tabular}\right),\qquad
\ket{\tilde{h}_2}=\frac{1}{\sqrt{\langle\tilde{h}_2\ket{\tilde{h}_2}}}
\left(\begin{tabular}{c} 
	0\\0\\$\lambda_2^2-(c^2+d^2)$ \\ $bc+ad e^{ik}$
\end{tabular}\right).
\eeq
It is obvious that these will also be eigenvectors of $\Gamma$.

The eigenvectors of the Hamiltonian, $\ket{\psi_{\pm l}}$ are  also eigenvectors of $H^2$, with eigenvalue $\lambda_l^2$. Therefore, for each value of $l$, we may write them as a normalized superposition of the two eigenvectors of $H^2$ with eigenvalue $\lambda_{l}^2$:
\beq
\ket{\psi_{\pm l}}=\hat{\alpha}_{\pm l}\ket{\hat{h}_l}+\tilde{\alpha}_{\pm l}\ket{\tilde{h}_l}.
\eeq
In particular, chiral symmetry imposes that energy eigenstates have equal support on both sublattices, i.e., $|\hat{\alpha}_{\pm l}|=|\tilde{\alpha}_{\pm l}|=1/\sqrt2$. Then, with an appropriate choice of phases, we can write them as:
\beq
\ket{\psi_{\pm l}}=\frac{\ket{\hat{h}_l}\pm e^{i \phi_{l}}\ket{\tilde{h}_l}}{\sqrt{2}}.
\eeq
The phase $\phi_{l}$ needs to be fixed imposing that $\ket{\psi_{\pm l}}$ is an eigenstate of $H$ with positive/negative energy. This may be done using the first line of the matrix equality $H\ket{\psi_{+l}}=+\lambda_l\ket{\psi_{+l}}$, which yields:
\beq
e^{i \phi_{l}}=\frac{\lambda_l\ket{\hat{h}_l}_1}{a\ket{\tilde{h}_l}_3+d e^{-ik}\ket{\tilde{h}_l}_4},
\eeq
where $\ket{\psi}_n$ indicates the $n^{\rm th}$ component of the vector $\ket{\psi}$.
Note finally that, upon sending $k\rightarrow -k$, the eigenstates of $H$ satisfy
\beq
\ket{\psi_{\pm l}(-k)}=\ket{\psi_{\pm l}(k)}^*,
\eeq
which tells us that the Hamiltonian is time-reversal symmetric. 
Now we can explicitly build the $Q$-matrix in the canonical chiral eigenbasis, it reads:
\beq
Q=\sum_{l=1,2}e^{i \phi_{l}}\ket{\tilde{h}_l}\bra{\hat{h}_l}+e^{-i \phi_{l}}\ket{\hat{h}_l}\bra{\tilde{h}_l} 
= \sum_{1\leq r,s,t\leq 4} \ket{\Gamma_r} M_{rs} \Gamma_{ss} (M^\dagger)_{st}\bra{\Gamma_t},
\eeq
with $M_{rs}=\bra{\Gamma_r}\psi_s\rangle$ the unitary matrix for the change of basis between the canonical-chiral and energy eigenstates. Computing the determinant of $q$, the lower-left block of $Q$, we see that ${\rm arg} [{\rm Det}(q)]=- i {\rm log}\left(\frac{ac-bde^{ik}}{|ac-bde^{ik}|}\right)={\rm arg}{[{\rm Det}(h)]}$. The winding of the SSH$_4$ model may now be computed from Eq.\ \eqref{gammaAO}, or equivalently from Eq.\ \eqref{gammaR}.

%

\end{document}